\documentclass[a4paper]{article}

\usepackage{INTERSPEECH2022}
\usepackage{amsmath}
\usepackage{multirow}
\usepackage{graphicx}
\usepackage{amssymb}
\usepackage{pifont}
\newcommand{\cmark}{\ding{51}}%
\newcommand{\xmark}{\ding{55}}%
\usepackage{hyperref}

\title{Improved Relation Networks for End-to-End Speaker Verification and Identification}
\name{Ashutosh Chaubey, Sparsh Sinha, Susmita Ghose}
\address{Data Science, LG Ads Solutions, Mountain View, California, USA}
\email{ashutoshchaubey@lgads.tv, sparshsinha@alphonso.tv, susmita@alphonso.tv}

\begin{document}

\maketitle

\begin{abstract}
  Speaker identification systems in a real-world scenario are tasked to identify a speaker amongst a set of enrolled speakers given just a few samples for each enrolled speaker. This paper demonstrates the effectiveness of meta-learning and relation networks for this use case. We propose improved relation networks for speaker verification and few-shot (unseen) speaker identification. The use of relation networks facilitates joint training of the frontend speaker encoder and the backend model. Inspired by the use of prototypical networks in speaker verification and to increase the discriminability of the speaker embeddings, we train the model to classify samples in the current episode amongst all speakers present in the training set. Furthermore, we propose a new training regime for faster model convergence by extracting more information from a given meta-learning episode with negligible extra computation. We evaluate the proposed techniques on VoxCeleb, SITW and VCTK datasets on the tasks of speaker verification and unseen speaker identification. The proposed approach outperforms the existing approaches consistently on both tasks.
  
\end{abstract}

\section{Introduction}
Speaker identification and verification systems are widely used in practical settings for authentication, security, and personalized recommendations. The goal of speaker identification is to identify the speaker present in a given test sample amongst the enrolled speakers, given a few voice samples for each speaker in the enrollment set. Speaker verification models, on the other hand, try to classify whether speakers in a given pair of test clips are the same. For speaker verification, traditionally, a frontend model such as i-vector \cite{dehak2011ivector, rohdin2018ivector} is used to extract speaker embeddings. A backend model is then used to compare embeddings from different voice samples \cite{prince2007plda, GarciaRomero2011AnalysisOI}. Embeddings corresponding to each enrolled speaker are compared with the test sample embeddings for speaker identification. The speaker with maximum similarity is identified as the speaker in the test sample.

In recent years, deep learning based speaker verification systems have gained traction, achieving superior performance compared to the traditional methods \cite{ bai2020deeplearningspeakerrecognitionsurvey}. Most of these networks are trained on speaker classification tasks with additional objectives to reduce intra-class variance and increase inter-class separability \cite{jung2019RawNet}. For final classification, a naive cosine similarity metric is typically used and works well in most cases. Several backend classification models \cite{lee2014bvector, ferrer2022robustspeakerverificationbackend} have also been proposed for this purpose, but most of these models are not trained end-to-end with the frontend.

More recently, meta-learning based approaches have proven to be very efficacious for learning good speaker embeddings for identification and verification \cite{Kumar2020DesigningNSmetalearningsurvey}. Prototypical network loss for speaker verification outperformed the state-of-the-art triplet loss \cite{wang2019protocentroid, ko2020protospeakerverification}. Prototypical networks have also been used for short utterance speaker verification, with an additional global classification loss for classifying samples present within an episode amongst all speakers present in the training data \cite{Kye2020MetaLearningFS}. Additional contrastive loss and transformation coefficients were introduced to improve the quality of learned speaker embeddings \cite{Chen2021Improvedmetalearningforspeakerverification}. These techniques use meta-learning for learning only a  discriminative frontend for extracting speaker embeddings.

Inspired by the effectiveness of relation networks in few-shot image classification tasks \cite{sung2018relationnetworksmetalearning}, we propose relation networks for speaker identification and verification. Instead of using cosine similarity to compare two speaker embeddings in a meta-learning episode, we propose using a \textit{relation network} that computes the similarity between two speaker embeddings. Additionally, we improve upon the input to the relation network inspired by the b-vector approach \cite{lee2014bvector}. Inspired by \cite{Kye2020MetaLearningFS}, we include extra supervision to the model in terms of global classification to learn discriminative embeddings.

Meta-learning-based speaker recognition techniques converge slower than vanilla speaker classification-based techniques \cite{Desplanques2020ECAPATDNNEC}. To improve that, we introduce a novel yet simple training regime that utilizes information from the samples present in a single meta-learning episode more efficiently.

In summary, major contributions of this work are (\emph{i}) improved relation network-based model for speaker verification and identification which can be trained end-to-end; \emph{(ii)} introduction of global classification with relation networks, in addition to local classification, for better supervision during training; and \emph{(iii)} a new training regime which leads to faster convergence. To the best of our knowledge, this work is the first detailed analysis of the effectiveness of relation networks for speaker verification. Furthermore, we report the performance of our approach on VoxCeleb \cite{nagrani2017voxceleb1, chung2018voxceleb2}, SITW \cite{McLaren2016sitw} and VCTK \cite{Veaux2016vctk} datasets and demonstrate better performance.

The rest of this paper is organized as follows. Section \ref{sec:related_works} introduces meta-learning approaches for speaker verification and relation networks. Section \ref{sec:method} describes the proposed approach in detail. In Section \ref{sec:experiments}, we detail the experimental setup and report the results. Finally, we provide a conclusion to this work in Section \ref{sec:conclusion}.

\section{Related works}
\label{sec:related_works}
\subsection{Meta-learning}
Meta-learning is one of the most common approaches for few-shot learning tasks \cite{Hospedales2021MetaLearningSurvey}. In an $N$-way $k$-shot classification scenario, training occurs episodically. Each mini-batch consists of a \emph{support} (enrollment) set $S$ with $k$ labelled samples for each of the $N$ classes, and an unlabelled \emph{query} (test) set $Q$. $(x^s_{c, i}$, $y^s_{c, i})$ represents the $i^{th}$ support sample for class $c$ and $(x^q_{j}, y^q_{j}$) represents the $j^{th}$ query sample (here $i \in \{1,...,k\}$; $j \in \{1,...,|Q|\}$; $c, y^s_{c, i}, y^q_{j} \in \{1,...,N\}$).

Prototypical networks \cite{wang2019protocentroid, snell2017prototypicalnetworks} for speaker verification use a speaker encoder (frontend) $f_\theta$ to encode each of the samples present in the current episode to an $M$-dimensional speaker embedding. For each of the $N$ speaker classes present in the current episode, prototypes $\nu_c$ are computed by averaging the speaker embeddings of the $k$ support samples for class $c$. The probability distribution $P$ over speaker classes for a given query sample $x^q$ is written as a softmax over the similarities $\mathcal{S}$ between the speaker embedding $f_\theta(x^q)$ of the test sample with all the prototypes of $N$ speaker classes. 
\begin{equation}
  P(y^q = c | x^q) = \frac{exp(\mathcal{S}(f_\theta(x^q), \nu_c))}{\sum_{c' = 1}^{N} exp(\mathcal{S}(f_\theta(x^q), \nu_{c'}))}
  \label{eq:proto_local}
\end{equation}
where $y^q \in \{1,...,N\}$ is the label for sample $x^q$.




Relation networks \cite{sung2018relationnetworksmetalearning} for few-shot learning use a relation network $g_\phi$ (backend) for computing the \emph{relation} $r$ between two embeddings. Given embeddings $f_\theta(x_i)$ and $f_\theta(x_j)$ corresponding to a pair of samples, the relation network provides a similarity score or \emph{relation} $r_{i,j} \in [0, 1]$ for that pair given by,  
\begin{equation}
  r_{i, j} = g_\phi(\mathcal{C}[f_\theta(x_i), f_\theta(x_j)])
  \label{eq:relation_vanilla}
\end{equation}
where $\mathcal{C}[., .]$ represents a concatenation of the embeddings channel-wise. A relation network thus provides a scope for learning the similarity between a pair of embeddings.

\subsection{DNN-based speaker encoder}

\begin{table}[!t]
\caption{Summary of the speaker encoder network. $T$: temporal dimension of feature maps, $s_k$: kernel size, $s_t$: stride, $d$: dilation, $n_f$: number of features.}
\label{tab:speaker_encoder}
\centering
\begin{tabular}{c|c|c|c}
\hline \hline
\textbf{\begin{tabular}[c]{@{}c@{}}Layer/\\Block\end{tabular}} & \textbf{Composition}                                                                              & \textbf{\begin{tabular}[c]{@{}c@{}}Input \\ size\end{tabular}} & \textbf{\begin{tabular}[c]{@{}c@{}}Output \\ size\end{tabular}} \\ \hline \hline
conv1                                                             & \begin{tabular}[c]{@{}c@{}}1D Convolution \\$s_k = 5$; $s_t = 2$\end{tabular}             & 80 x T                                                         & 1024 x T                                                        \\ \hline
conv2                                                             & \begin{tabular}[c]{@{}c@{}}Dilated SE Res2Block\\ $s_k = 3$; $d = 2$\end{tabular}          & 1024 x T                                                       & 1024 x T                                                        \\ \hline
conv3                                                             & \begin{tabular}[c]{@{}c@{}}Dilated SE Res2Block\\ $s_k = 3$; $d = 3$\end{tabular}          & 1024 x T                                                       & 1024 x T                                                        \\ \hline
conv4                                                             & \begin{tabular}[c]{@{}c@{}}Dilated SE Res2Block\\ $s_k = 3$; $d = 4$\end{tabular}          & 1024 x T                                                       & 1024 x T                                                        \\ \hline
conv5                                                             & \begin{tabular}[c]{@{}c@{}}1D Convolution\\ $s_k = 1$; $s_t = 2$\end{tabular}             & 3072 x T                                                       & 1536 x T                                                        \\ \hline
atten                                                         & \begin{tabular}[c]{@{}c@{}}Attentive Stat. Pool. \cite{Desplanques2020ECAPATDNNEC}\end{tabular} & 1536 x T                                                       & 3072 x 1                                                        \\ \hline
FC                                                                & \begin{tabular}[c]{@{}c@{}} FC; $n_f = 192$\end{tabular}                         & 3072 x 1                                                       & 192 x 1                \\  \hline \hline                               
\end{tabular}
\end{table}
Inspired from \cite{Desplanques2020ECAPATDNNEC}, we use a TDNN-based speaker encoder with channel- and context-dependent statistics pooling and 1D Squeeze Excitation (SE) Res2Blocks. Table \ref{tab:speaker_encoder} summarizes the architecture of the speaker encoder. The first layer is a 1D convolutional layer with a kernel size of 5 and a stride of 2 with 1024 channels. 

The first layer is followed by three subsequent dilated SE Res2Blocks \cite{gao2021res2net}. Each dilated SE Res2Block consists of a Res2Block \cite{gao2021res2net} preceded and followed by a 1D convolutional layer with kernel size one. Finally, there is a squeeze excitation (SE) block which introduces channel interdependencies within each dilated SE Res2Block \cite{hu2018senet}. All the layers involved in the dilated SE Res2Blocks have 1024 channels. The multi-level features obtained from the three dilated SE Res2Blocks are then aggregated through a 1D convolutional layer. 

The frame-level embeddings are aggregated using channel- and context-dependent attentive statistics pooling \cite{Desplanques2020ECAPATDNNEC} to include attention to both channel and temporal dimensions of the feature map. The aggregated feature is then passed through a fully connected layer to get the utterance-level speaker embedding.

\section{Method}
\label{sec:method}
\begin{figure}[!t]
  \centering
  \includegraphics[width=\linewidth]{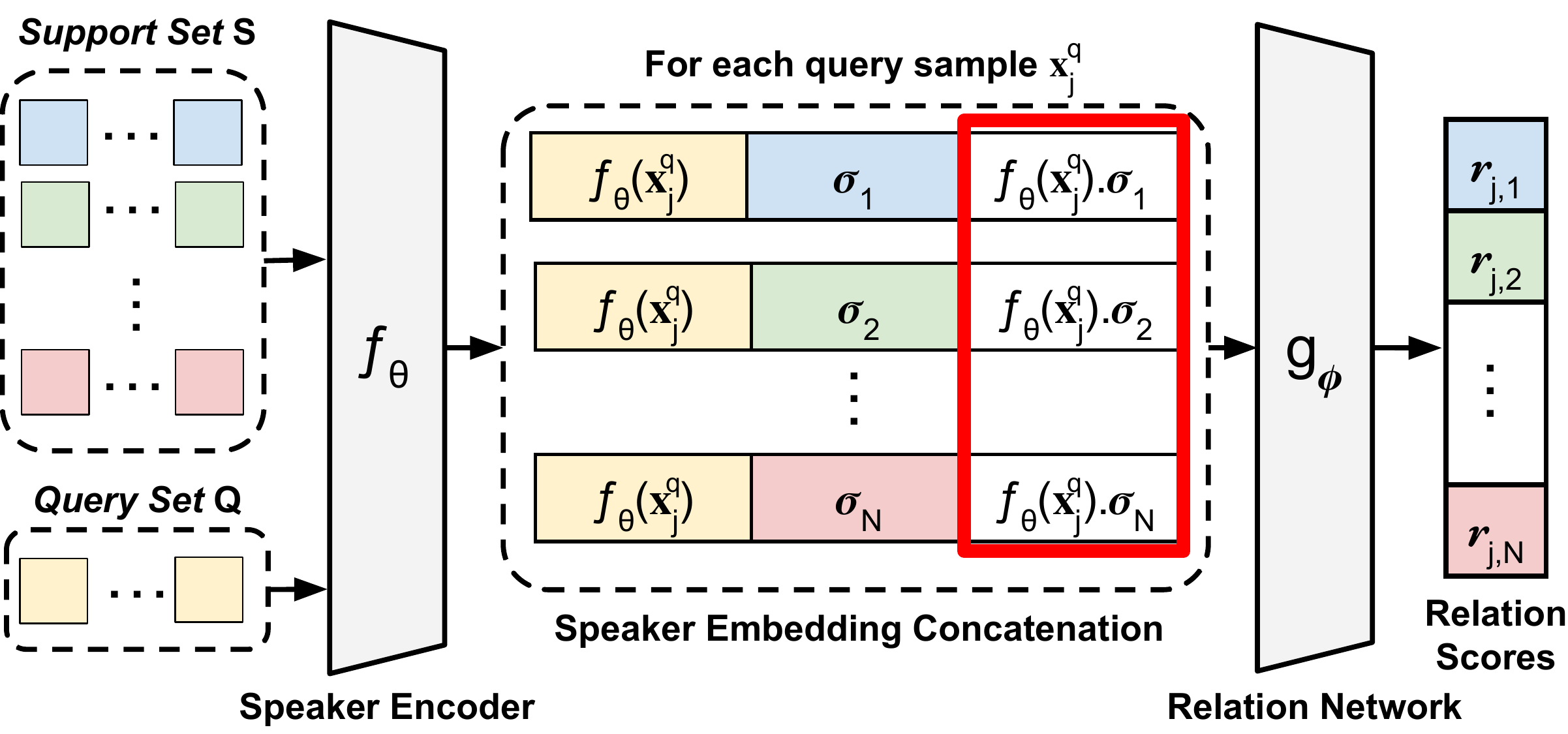}
  \caption{Schematic diagram of the proposed relation network based approach. Red box indicates the additional element-wise multiplication term input to the relation network. Colours in the support set represent different classe $c \in \{1,...,N\}$.}
  \label{fig:relation_net}
\end{figure}
\subsection{Improved relation networks}
In this work, given a query sample $x^q_j$ in an episode, we propose to compute the relation score between $x^q_j$ and support samples $x^s_{c, i}$ for each of the $c \in \{1,...,N\}$ classes as depicted in Figure \ref{fig:relation_net}. First, we pass each sample present in an episode through the speaker encoder $f_\theta$ to get the corresponding speaker embeddings. For each class, we average out the embeddings of the support samples to generate an aggregated representation $\sigma_c$ of that speaker.
\begin{equation}
  \sigma_c = \frac{1}{|S_c|} \sum_{i=1}^{|S_c|} f_\theta(x^s_{c, i})
  \label{eq:sigma_c_computation}
\end{equation}

For computing the relation $r_{j,c}$ between embedding pair $f_\theta(x^q_j)$ and $\sigma_c$, instead of just concatenating $f_\theta(x^q_j)$ and $\sigma_c$ as in Eqn. (\ref{eq:relation_vanilla}), we propose an additional concatenation of element-wise multiplication of the two speaker embeddings (red box in Figure \ref{fig:relation_net}), i.e. 
\begin{equation}
  r_{j, c} = g_\phi(\mathcal{C}[f_\theta(x^q_j), \sigma_c, f_\theta(x^q_j) \cdot \sigma_c])
  \label{eq:relation_bvector}
\end{equation}
where $\mathcal{C}[., ., .]$ represents a concatenation of the embeddings channel-wise. Using this additional multiplication operation helps in identifying whether the embeddings correspond to the same speaker or not \cite{jung2019RawNet, lee2014bvector}. 

As suggested in \cite{sung2018relationnetworksmetalearning}, we use mean square error (MSE) as the local loss $\mathcal{L}_{local}$ for training our model, regressing the score $r_{j,c}$ to the ground truth, i.e. 
\begin{equation}
  \mathcal{L}_{local} = \sum_{j=1}^{|Q|}\sum_{c=1}^{N} (r_{j, c} - \mathbf{1}(y^q_j == c))
  \label{eq:loss_local}
\end{equation}
where $y^q_j \in \{1,...,N\}$ represents the ground truth for the sample $x^q_j$, and $\mathbf{1}(.)$ is 1 if the condition inside bracket is true, else 0.

\subsection{Global classification}
\label{subsec:global_classification}
Inspired from \cite{Kye2020MetaLearningFS}, we provide extra supervision by using global classification. We assume $\mathcal{W} = \{\omega_C| C \in \{1,...,N'\}\}$ as representations or \emph{prototypes} for all the $N'$ classes present in the entire dataset. Then the relation $r_{j,C}$ between the query sample $x^q_j$ and representation $\omega_C$ is,
\begin{equation}
  r_{j, C} = g_\phi(\mathcal{C}[f_\theta(x^q_j), \omega_C, f_\theta(x^q_j) \cdot \omega_C])
  \label{eq:relation_global}
\end{equation}

The corresponding global loss function $\mathcal{L}_{global}$
\begin{multline}
    \mathcal{L}_{global} = \sum_{j=1}^{|Q|}\sum_{C=1}^{N'} (r_{j, C} - \mathbf{1}(Y^q_j == C)) \\+ \sum_{i=1}^{|S|}\sum_{C=1}^{N'} (r_{i, C} - \mathbf{1}(Y^s_i == C))
  \label{eq:loss_global}
\end{multline}
where $Y^s_i, Y^q_j \in \{1,...,N'\}$ are the global labels for the sample $x^s_i$ and $x^q_j$. Note that here, we are computing the loss for both the support and query samples.

The relation loss function $\mathcal{L}_{total}$ including the global loss is given by,
\begin{equation}
\label{eq:total_loss}
    \mathcal{L}_{total} = \mathcal{L}_{local} + \lambda * \mathcal{L}_{global}
\end{equation}
where $\lambda$ is a hyperparameter.

Training the relation network happens in stages. First, we train just using $\mathcal{L}_{local}$ (Eqn. (\ref{eq:loss_local})), optimizing the relation network through episodic supervision. Then, $\omega_C$ for all the classes in the training set is initialized as the average of speaker embeddings $f_\theta(x_{C, i})$ for all the samples belonging to that class in the training set. The network is then fine-tuned with added global supervision with the objective mentioned in Eqn. (\ref{eq:total_loss}).

\subsection{Improved training regime}
\label{subsec:improved_regime}
\begin{figure}[!ht]
  \centering
  \includegraphics[width=\linewidth]{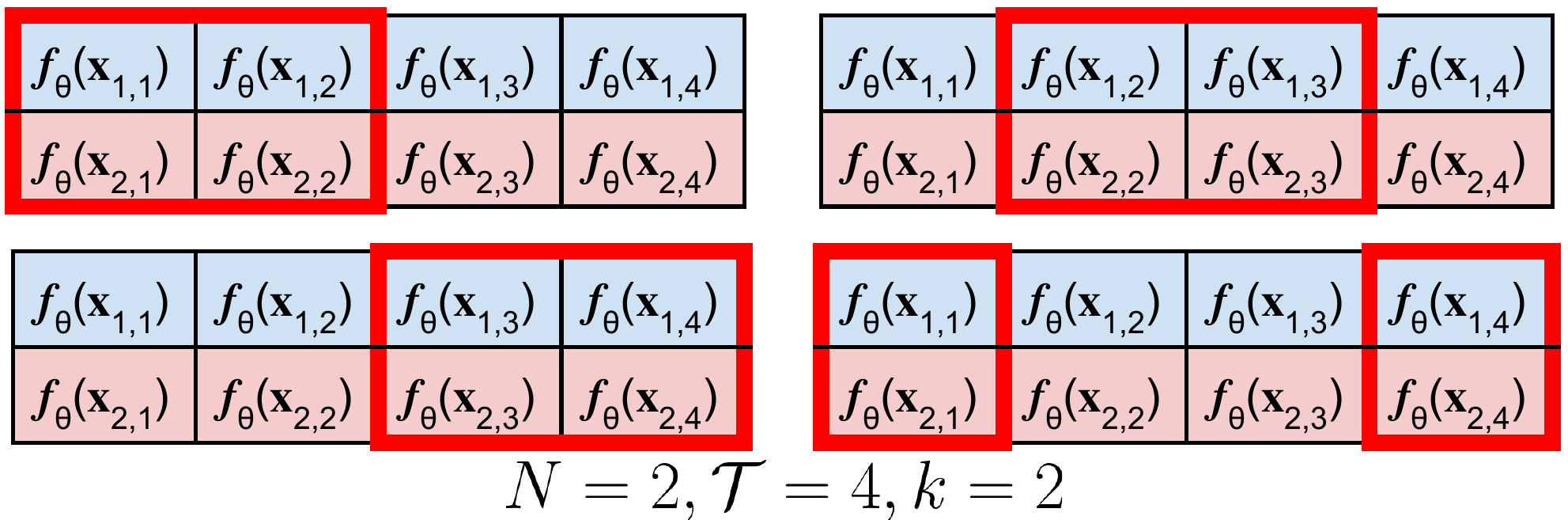}
  \caption{Different support-query combinations. The red box represents the support set, and the rest of the embeddings are in the query set. Colours represent different classes. Instead of using the first $k$ embeddings from each class as the support set, we use $k$ consecutive embeddings in cyclic order.}
  \label{fig:improved_regime}
\end{figure}

For a class $c$ present in the current episode, the support samples $x^s_{c, i}$ and query samples $x^q_{c, j}$ are chosen randomly and can be used interchangeably. Since obtaining the speaker embeddings through the speaker encoder $f_\theta$ is the bottleneck for computation in the training pipeline, we propose using samples present in the current episode by creating multiple support-query combinations as depicted in Figure \ref{fig:improved_regime}.

For a given episode, we create $\mathcal{T}$ support-query combinations, where $\mathcal{T}$ is the total number of samples for each class within that episode. Given speaker embeddings $F_c = \{f_\theta(x_{c,1}),...,f_\theta(x_{c,\mathcal{T}})\}$ for each of the $N$ classes in the current episode, we propose creating multiple support-query combinations $X_{s,q}$ as follows,
\begin{equation}
    X_{s,q} = \bigcup_{l=1}^{\mathcal{T}} \{(S_l = \{S_{l, 1},...,S_{l,N}\}, Q_l = \{Q_{l, 1},...,Q_{l,N}\})\}
    \label{eq:improved_regime}
\end{equation}
\begin{equation}
    S_{l,c} = \{\mathcal{H}(c,l),...,\mathcal{H}(c,l+(k-1))\}
    \label{eq:s_lc}
\end{equation}
\begin{equation}
    Q_{l,c} = \{\mathcal{H}(c,l+k),...,\mathcal{H}(c,l+(\mathcal{T}-1))\}
    \label{eq:q_lc}
\end{equation}
\begin{equation}
    \mathcal{H}(c, z) = f_\theta(x_{c,1+(z-1)\%\mathcal{T}})
    \label{eq:h_cz}
\end{equation}

where ($S_l$,$Q_l$) is the $l^{th}$ support-query combination, ($S_{l,c}$, $Q_{l, c}$) are the support and query samples for a class $c$ in the $l^{th}$ support-query combination, $c \in \{1,...,N\}$. $\%$ denotes a modulo operation and $\mathcal{H}(c,z)$ is the cyclic index function with the property $\mathcal{H}(c, 1) = f_\theta(x_{c,1})$,...,$\mathcal{H}(c, \mathcal{T}) = f_\theta(x_{c,\mathcal{T}})$, $\mathcal{H}(c, \mathcal{T}+1) = f_\theta(x_{c,1})$,...,$\mathcal{H}(c, 2\mathcal{T}) = f_\theta(x_{c,\mathcal{T}})$ and so on. 

Losses for each of the support-query combinations within an episode are aggregated and backpropagated together, only once. Note that there are more such support-query combinations possible but we only use $\mathcal{T}$ such pairs obtained from Eqn (\ref{eq:improved_regime}) as above for computational simplicity.

\section{Experiments}
\label{sec:experiments}
\subsection{Datasets}

We use the VoxCeleb1 \cite{nagrani2017voxceleb1}, SITW \cite{McLaren2016sitw} and VCTK \cite{Veaux2016vctk} datasets to evaluate the proposed methods on the tasks of speaker verification and few-shot (unseen) speaker identification. We use the dev set of VoxCeleb2 \cite{chung2018voxceleb2} for training all the model variations. Since neural networks benefit from data augmentation, four different types of augmentations are added to the training clips following the Kaldi recipe \cite{Povey2011kaldi} in combination with the MUSAN (music, babble, noise) \cite{Snyder2015MUSANAM} and RIR (reverb) \cite{ko2017rir} dataset. 

For speaker verification, we use the standard trial pairs for the VoxCeleb1 test set \cite{nagrani2017voxceleb1} and eval \emph{core-core} trial pairs for the SITW dataset \cite{McLaren2016sitw} for evaluating our model. We use Equal Error Rate (EER) and the minimum detection cost function (minDCF or $C_{det}^{min}$) with $P_{target} = 0.01$ and $C_{FA} = C_{Miss} = 1$ for comparing the proposed methods with the baselines. For unseen speaker identification, we evaluate the model using the VCTK corpus \cite{Veaux2016vctk}. We run 1000 randomly generated episodes and report the average accuracy with 95\% confidence intervals. 

\subsection{Experimental setup}
We use 80-dimensional log mel-filterbanks with a window size of 25ms and a hop of 10ms as input features. Input features are mean-normalized across the time axis, and no voice activity detection (VAD) is applied. As an additional augmentation step, SpecAugment \cite{park2019specaug} is applied by randomly masking 0-10 frames in the time axis and 0-8 frames in the frequency axis. 

All variations in the models along with the baseline models have been trained with an initial learning rate of 0.001 with a step-wise decay of 0.97 every epoch. We use Adam optimizer with $\beta = \{0.9, 0.999\}$ and a weight-decay of 2e-5. The meta-learning-based approaches have been trained by keeping 120 speakers in each episode with one support sample and two query samples for each speaker. For a fair comparison, we train the vanilla methods with a batch size of 360. All the relation networks are fully connected networks. We use leaky ReLU activations for better gradient flow and dropout in between the layers of the relation network to avoid overfitting. A single NVIDIA GeForce RTX 3090 GPU was used to run all our experiments.

\subsection{Speaker verification}

\begin{table}[]
\centering
\caption{Performance of various systems for speaker verification. S: softmax, AAM: AAM-softmax \cite{deng2019aamsoftmax}, M: meta-learning.}
\label{tab:ver_id_results}
\resizebox{\linewidth}{!}{
\begin{tabular}{ccc||cc||cc}
\hline \hline
\multicolumn{3}{c||}{\textbf{System Configuration}}                                                                                                                                               & \multicolumn{2}{c||}{\textbf{VoxCeleb1}}               & \multicolumn{2}{c}{\textbf{SITW}}                    \\ \hline 
\multicolumn{1}{c|}{Model + Loss}                                                     & \multicolumn{1}{c|}{Backend}                                          & M & \multicolumn{1}{c|}{EER\%} & \textbf{$C_{det}^{min}$} & \multicolumn{1}{c|}{EER\%} & \textbf{$C_{det}^{min}$} \\ \hline \hline
\multicolumn{1}{c|}{TDNN + S}                                                                  & \multicolumn{1}{c|}{cosine}                                                     & \xmark             & \multicolumn{1}{c|}{2.466}                & 0.2623             & \multicolumn{1}{c|}{4.733}                & 0.6858             \\
\multicolumn{1}{c|}{TDNN + AAM}                                                                & \multicolumn{1}{c|}{cosine}                                                     & \xmark             & \multicolumn{1}{c|}{1.262}                & \textbf{0.1517}             & \multicolumn{1}{c|}{2.735}                & 0.3633             \\
\multicolumn{1}{c|}{TDNN + AAM}                                                                & \multicolumn{1}{c|}{\begin{tabular}[c]{@{}c@{}}Relation\\Net\end{tabular}}                                                   & \xmark             & \multicolumn{1}{c|}{1.452}                & 0.2484             & \multicolumn{1}{c|}{3.235}                & 0.4139             \\ \hline
\multicolumn{1}{c|}{\begin{tabular}[c]{@{}c@{}}TDNN + \\ Prototypical \cite{Kye2020MetaLearningFS}\end{tabular}}                                                       & \multicolumn{1}{c|}{cosine}                                                     & \cmark             & \multicolumn{1}{c|}{1.235}                & 0.1961             & \multicolumn{1}{c|}{2.277}                & 0.3934             \\
\multicolumn{1}{c|}{\begin{tabular}[c]{@{}c@{}}TDNN + Relation\\ Vanilla (\textbf{Ours})\end{tabular}}  & \multicolumn{1}{c|}{\begin{tabular}[c]{@{}c@{}}Relation\\Net\end{tabular}} & \cmark             & \multicolumn{1}{c|}{1.225}                & 0.1548             & \multicolumn{1}{c|}{2.237}                & 0.3328             \\
\multicolumn{1}{c|}{\begin{tabular}[c]{@{}c@{}}TDNN + Relation\\ Improved (\textbf{Ours})\end{tabular}} & \multicolumn{1}{c|}{\begin{tabular}[c]{@{}c@{}}Relation\\Net\end{tabular}} & \cmark             & \multicolumn{1}{c|}{\textbf{1.193}}                & 0.1523             & \multicolumn{1}{c|}{\textbf{2.050}}                & \textbf{0.3187}             \\ \hline \hline
\end{tabular}}
\end{table}

Table \ref{tab:ver_id_results} summarizes the results of the proposed approach on speaker verification. We report the performances for two variations of our relation network-based model. The first variation uses relation as described in Eqn. (\ref{eq:relation_vanilla}) while the other uses the additional multiplicative term described in Eqn. (\ref{eq:relation_bvector}). We can see that the improved relation network performs better than the other methods on both VoxCeleb1 and SITW. Meta-learning based approaches in general show superior performance as compared to the speaker classification-based approaches. A backend with architecture same as relation network (Eqn. (\ref{eq:relation_bvector})) was trained with the embeddings of TDNN + AAM frontend separately. Better performance over this setting indicates that the proposed end-to-end meta-learning is better than training the frontend and backend models separately. Moreover, the superior performance of the relation network-based approaches over prototypical networks highlight the importance of increased flexibility in terms of having a learnable backend.


\subsection{Global classification}

\begin{table}[]
\centering
\caption{Performance of the proposed approach with added global supervision. $\lambda$: weight hyperparameter for global classification in Eqn. (\ref{eq:total_loss}).}
\label{tab:global_classification}
\resizebox{\linewidth}{!}{%
\begin{tabular}{cc||cc||cc} 
\hline \hline
\multicolumn{2}{c||}{\textbf{System Configuration}} & \multicolumn{2}{c||}{\textbf{VoxCeleb1}} & \multicolumn{2}{c}{\textbf{SITW}} \\ \hline
\multicolumn{1}{c|}{Model}              & $\lambda$   & \multicolumn{1}{c|}{EER\%}    & $C_{det}^{min}$    & \multicolumn{1}{c|}{EER\%} & $C_{det}^{min}$ \\ \hline \hline
\multicolumn{1}{c|}{TDNN + Relation}    & 0        & \multicolumn{1}{c|}{1.543}         &    0.2157     & \multicolumn{1}{c|}{2.570}      &   0.4744      \\
\multicolumn{1}{c|}{TDNN + Relation}    & 0.5      & \multicolumn{1}{c|}{1.410}         &    0.1681    & \multicolumn{1}{c|}{2.214}      &    0.4168  \\
\multicolumn{1}{c|}{TDNN + Relation}    & 1.0      & \multicolumn{1}{c|}{1.336}         &    0.1688     & \multicolumn{1}{c|}{2.324}      &   0.4103  \\ \hline\hline
\end{tabular}%
}
\end{table}

To evaluate and prove the efficacy of the added global supervision introduced in Section \ref{subsec:global_classification}, we trained the proposed approach by varying $\lambda$ in Eqn (\ref{eq:total_loss}) and report the performance of the trained models in Table \ref{tab:global_classification}. All the model variations are trained on VoxCeleb2. Each episode has 30 speakers having one support and one query sample. As we increase the value of $\lambda$, we see a significant increase in the model's performance. This gain in performance can be attributed to the additional supervision that is provided by global classification. These results validate that global classification improves the overall discriminability of the model. 

\subsection{Improved training regime}
\begin{figure}[]
  \centering
  \includegraphics[trim={0 0.5cm 0 2.9cm},clip,width=0.9\linewidth]{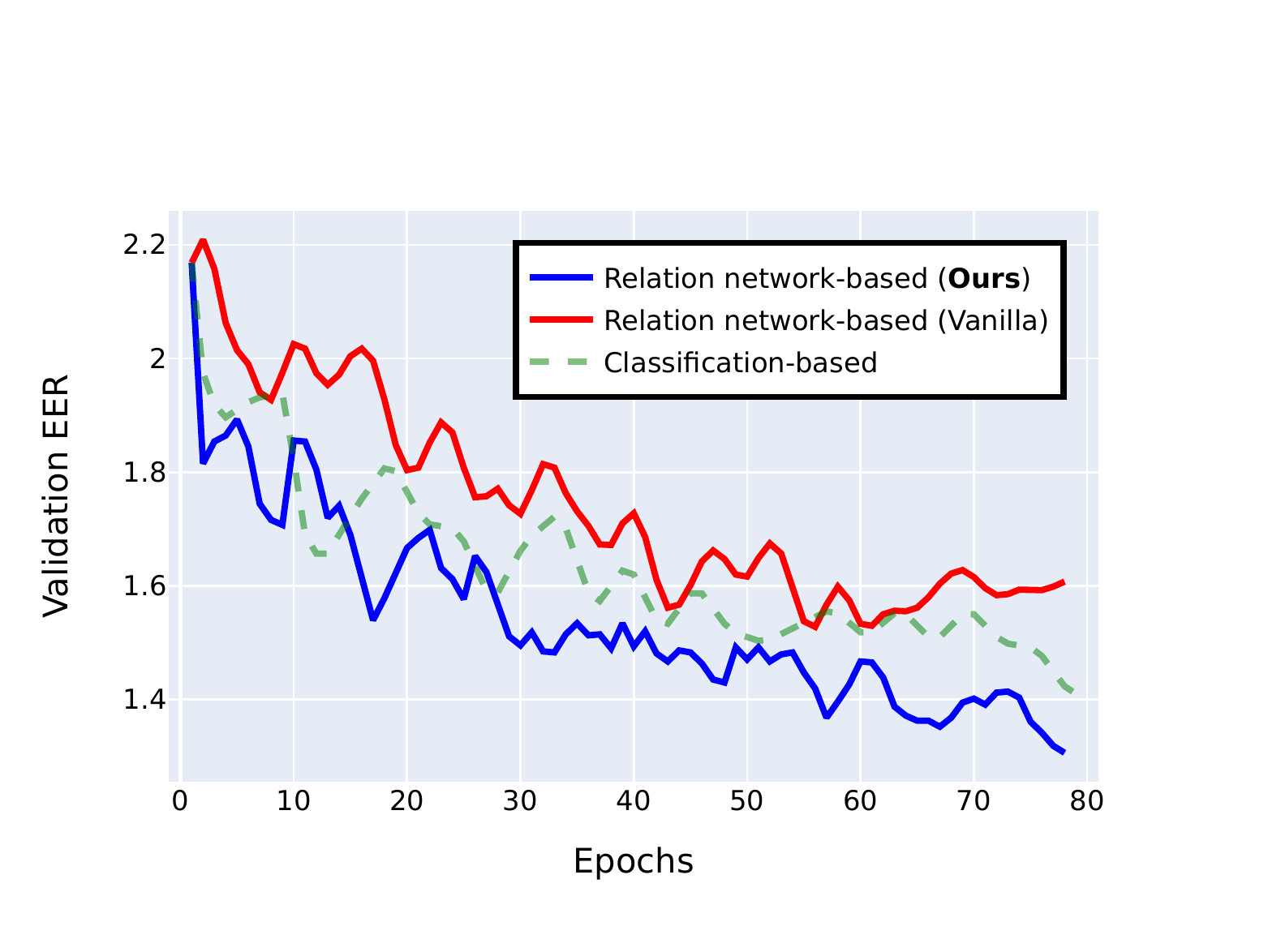}
  \caption{Training curves for different training techniques. Avg. training time per epoch - Relation network-based(Ours): 25.05 min, Relation network-based(Vanilla): 23.97 min, Classification based: 24.94 min.}
  \label{fig:improved_regime_plot}
\end{figure}

To analyze our training technique, we trained the relation network model on VoxCeleb2 from the same starting point using three different approaches. The vanilla approach trains the relation network-based model using the embeddings of samples in the current episode just once, while our approach uses the embeddings more efficiently (refer Section \ref{subsec:improved_regime}). We also trained a classification-based TDNN + AAM model to compare the meta-learning-based approaches with it. Figure \ref{fig:improved_regime_plot} shows the training curve for the different approaches for 80 epochs. Validation EERs are reported on trial pairs constructed from a small subset ($2\%$) of VoxCeleb2 disjoint from the training set. We can observe that our improved training regime leads to faster convergence. 
Both the vanilla and our improved technique have almost equal computation times as the time to process the embeddings through the relation network is negligible compared to generating the embeddings from the speaker encoder.

\subsection{Unseen speaker identification}

Unseen speaker identification means that none of the speakers used in testing has been used while training the model. We evaluate the performance of unseen speaker identification on VCTK \cite{Veaux2016vctk} corpus with one support and five query samples in each test episode. As shown in Table \ref{tab:identification_results}, the proposed relation network-based approach performs better than the TDNN + Prototypical \cite{Kye2020MetaLearningFS} as well as TDNN + AAM \cite{Desplanques2020ECAPATDNNEC} model. Improvement in the performance using relation networks increases as the number of speakers increases. Identifying the correct speaker within the enrollment set becomes more difficult with an increase in the number of speakers.

\begin{table}[]
\centering
\caption{Unseen speaker identification performance on VCTK dataset. AAM: AAM-softmax \cite{deng2019aamsoftmax}, M: meta-learning.}
\label{tab:identification_results}
\resizebox{\linewidth}{!}{%
\begin{tabular}{ccc||ccc}
\hline \hline
\multicolumn{3}{c||}{\textbf{System Configuration}}                                                                                                                               & \multicolumn{3}{c}{\textbf{Accuracy (\%)}}                          \\ \hline
\multicolumn{1}{c|}{Model + Loss}                                                              & \multicolumn{1}{c|}{Backend}                                                & M & \multicolumn{1}{c|}{10-way} & \multicolumn{1}{c|}{50-way} & 100-way \\ \hline \hline
\multicolumn{1}{c|}{TDNN + AAM}                                                                & \multicolumn{1}{c|}{cosine}                                                 & \xmark & \multicolumn{1}{c|}{98.84\begin{tiny}$\pm$0.07\end{tiny}}      & \multicolumn{1}{c|}{96.10\begin{tiny}$\pm$0.05\end{tiny}}      & 93.98\begin{tiny}$\pm$0.03\end{tiny}       \\
\multicolumn{1}{c|}{\begin{tabular}[c]{@{}c@{}}TDNN + \\ Prototypical\cite{Kye2020MetaLearningFS}\end{tabular}}            & \multicolumn{1}{c|}{cosine}                                                 & \cmark & \multicolumn{1}{c|}{98.34\begin{tiny}$\pm$0.10\end{tiny}}      & \multicolumn{1}{c|}{95.18\begin{tiny}$\pm$0.08\end{tiny}}      &    92.75\begin{tiny}$\pm$0.07\end{tiny}    \\
\multicolumn{1}{c|}{\begin{tabular}[c]{@{}c@{}}TDNN + Relation\\ Improved (\textbf{Ours})\end{tabular}} & \multicolumn{1}{c|}{\begin{tabular}[c]{@{}c@{}}Relation\\ Net\end{tabular}} & \cmark & \multicolumn{1}{c|}{\textbf{98.98\begin{tiny}$\pm$0.05\end{tiny}}}      & \multicolumn{1}{c|}{\textbf{96.94\begin{tiny}$\pm$0.04\end{tiny}}}      & \textbf{95.24\begin{tiny}$\pm$0.04\end{tiny}}       \\ \hline \hline
\end{tabular}%
}
\end{table}

\section{Conclusion}
\label{sec:conclusion}
This paper proposed improved relation networks for speaker verification and identification. Additionally, we provided two enhancements to our technique -  global classification for better supervision and an improved training regime for better utilization of samples present in an episode. The proposed method outperforms the existing techniques on speaker verification and identification on VoxCeleb, SITW and VCTK datasets. 

\bibliographystyle{IEEEtran}

\bibliography{mybib}

\begin{thebibliography}{10}
\providecommand{\url}[1]{#1}
\csname url@samestyle\endcsname
\providecommand{\newblock}{\relax}
\providecommand{\bibinfo}[2]{#2}
\providecommand{\BIBentrySTDinterwordspacing}{\spaceskip=0pt\relax}
\providecommand{\BIBentryALTinterwordstretchfactor}{4}
\providecommand{\BIBentryALTinterwordspacing}{\spaceskip=\fontdimen2\font plus
\BIBentryALTinterwordstretchfactor\fontdimen3\font minus
  \fontdimen4\font\relax}
\providecommand{\BIBforeignlanguage}[2]{{%
\expandafter\ifx\csname l@#1\endcsname\relax
\typeout{** WARNING: IEEEtran.bst: No hyphenation pattern has been}%
\typeout{** loaded for the language `#1'. Using the pattern for}%
\typeout{** the default language instead.}%
\else
\language=\csname l@#1\endcsname
\fi
#2}}
\providecommand{\BIBdecl}{\relax}
\BIBdecl

\bibitem{dehak2011ivector}
N.~Dehak, P.~J. Kenny, R.~Dehak, P.~Dumouchel, and P.~Ouellet, ``Front-end
  factor analysis for speaker verification,'' \emph{IEEE Transactions on Audio,
  Speech, and Language Processing}, vol.~19, no.~4, pp. 788--798, 2011.

\bibitem{rohdin2018ivector}
J.~Rohdin, A.~Silnova, M.~Diez, O.~Plchot, P.~Matějka, and L.~Burget,
  ``End-to-end dnn based speaker recognition inspired by i-vector and plda,''
  in \emph{2018 IEEE International Conference on Acoustics, Speech and Signal
  Processing (ICASSP)}, 2018, pp. 4874--4878.

\bibitem{prince2007plda}
S.~J. Prince and J.~H. Elder, ``Probabilistic linear discriminant analysis for
  inferences about identity,'' in \emph{2007 IEEE 11th International Conference
  on Computer Vision}, 2007, pp. 1--8.

\bibitem{GarciaRomero2011AnalysisOI}
D.~Garcia-Romero and C.~Y. Espy-Wilson, ``Analysis of i-vector length
  normalization in speaker recognition systems,'' in \emph{INTERSPEECH}, 2011.

\bibitem{jung2019RawNet}
J.-w. Jung, H.-s. Heo, j.-h. Kim, H.-j. Shim, and H.-j. Yu, ``Rawnet: Advanced
  end-to-end deep neural network using raw waveforms for text-independent
  speaker verification,'' \emph{Proc. Interspeech 2019}, pp. 1268--1272, 2019.

\bibitem{lee2014bvector}
H.-S. Lee, Y.~Tso, Y.-F. Chang, H.-M. Wang, and S.-K. Jeng, ``Speaker
  verification using kernel-based binary classifiers with binary operation
  derived features,'' in \emph{2014 IEEE International Conference on Acoustics,
  Speech and Signal Processing (ICASSP)}, 2014, pp. 1660--1664.

\bibitem{Kumar2020DesigningNSmetalearningsurvey}
M.~Kumar, T.~Jin-Park, S.~L. Bishop, and S.~S. Narayanan, ``Designing neural
  speaker embeddings with meta learning,'' \emph{ArXiv}, vol. abs/2007.16196,
  2020.

\bibitem{wang2019protocentroid}
J.~Wang, K.-C. Wang, M.~T. Law, F.~Rudzicz, and M.~Brudno, ``Centroid-based
  deep metric learning for speaker recognition,'' in \emph{ICASSP 2019 - 2019
  IEEE International Conference on Acoustics, Speech and Signal Processing
  (ICASSP)}, 2019, pp. 3652--3656.

\bibitem{ko2020protospeakerverification}
T.~Ko, Y.~Chen, and Q.~Li, ``Prototypical networks for small footprint
  text-independent speaker verification,'' in \emph{ICASSP 2020 - 2020 IEEE
  International Conference on Acoustics, Speech and Signal Processing
  (ICASSP)}, 2020, pp. 6804--6808.

\bibitem{Kye2020MetaLearningFS}
S.~M. Kye, Y.~Jung, H.~Lee, S.~J. Hwang, and H.~Kim, ``Meta-learning for short
  utterance speaker recognition with imbalance length pairs,'' in
  \emph{INTERSPEECH}, 2020.

\bibitem{Chen2021Improvedmetalearningforspeakerverification}
Y.~Chen, W.~Guo, and B.~Gu, ``Improved meta-learning training for speaker
  verification,'' in \emph{INTERSPEECH}, 2021.

\bibitem{sung2018relationnetworksmetalearning}
F.~Sung, Y.~Yang, L.~Zhang, T.~Xiang, P.~H. Torr, and T.~M. Hospedales,
  ``Learning to compare: Relation network for few-shot learning,'' in
  \emph{Proceedings of the IEEE Conference on Computer Vision and Pattern
  Recognition}, 2018.

\bibitem{Desplanques2020ECAPATDNNEC}
B.~Desplanques, J.~Thienpondt, and K.~Demuynck, ``Ecapa-tdnn: Emphasized
  channel attention, propagation and aggregation in tdnn based speaker
  verification,'' in \emph{INTERSPEECH}, 2020.

\bibitem{nagrani2017voxceleb1}
\BIBentryALTinterwordspacing
A.~Nagrani, J.~S. Chung, and A.~Zisserman, ``Voxceleb: A large-scale speaker
  identification dataset,'' \emph{Interspeech 2017}, Aug 2017. [Online].
  Available: \url{http://dx.doi.org/10.21437/Interspeech.2017-950}
\BIBentrySTDinterwordspacing

\bibitem{chung2018voxceleb2}
\BIBentryALTinterwordspacing
J.~S. Chung, A.~Nagrani, and A.~Zisserman, ``Voxceleb2: Deep speaker
  recognition,'' \emph{Interspeech 2018}, Sep 2018. [Online]. Available:
  \url{http://dx.doi.org/10.21437/Interspeech.2018-1929}
\BIBentrySTDinterwordspacing

\bibitem{McLaren2016sitw}
M.~McLaren, L.~Ferrer, D.~Cast{\'a}n, and A.~D. Lawson, ``The speakers in the
  wild (sitw) speaker recognition database,'' in \emph{INTERSPEECH}, 2016.

\bibitem{Veaux2016vctk}
C.~Veaux, J.~Yamagishi, and K.~MacDonald, ``Superseded - cstr vctk corpus:
  English multi-speaker corpus for cstr voice cloning toolkit,'' 2016.

\bibitem{Hospedales2021MetaLearningSurvey}
T.~M. Hospedales, A.~Antoniou, P.~Micaelli, and A.~J. Storkey, ``Meta-learning
  in neural networks: A survey,'' \emph{IEEE transactions on pattern analysis
  and machine intelligence}, vol.~PP, 2021.

\bibitem{snell2017prototypicalnetworks}
J.~Snell, K.~Swersky, and R.~Zemel, ``Prototypical networks for few-shot
  learning,'' in \emph{Proceedings of the 31st International Conference on
  Neural Information Processing Systems}, ser. NIPS'17.\hskip 1em plus 0.5em
  minus 0.4em\relax Red Hook, NY, USA: Curran Associates Inc., 2017, p.
  4080–4090.

\bibitem{gao2021res2net}
\BIBentryALTinterwordspacing
S.-H. Gao, M.-M. Cheng, K.~Zhao, X.-Y. Zhang, M.-H. Yang, and P.~Torr,
  ``Res2net: A new multi-scale backbone architecture,'' \emph{IEEE Transactions
  on Pattern Analysis and Machine Intelligence}, vol.~43, no.~2, p. 652–662,
  Feb 2021. [Online]. Available:
  \url{http://dx.doi.org/10.1109/TPAMI.2019.2938758}
\BIBentrySTDinterwordspacing

\bibitem{hu2018senet}
J.~Hu, L.~Shen, and G.~Sun, ``Squeeze-and-excitation networks,'' in \emph{2018
  IEEE/CVF Conference on Computer Vision and Pattern Recognition}, 2018, pp.
  7132--7141.

\bibitem{Povey2011kaldi}
D.~Povey, A.~K. Ghoshal, G.~Boulianne, L.~Burget, O.~Glembek, N.~K. Goel,
  M.~Hannemann, P.~Motl{\'i}cek, Y.~Qian, P.~Schwarz, J.~Silovsk{\'y},
  G.~Stemmer, and K.~Vesel{\'y}, ``The kaldi speech recognition toolkit,''
  2011.

\bibitem{Snyder2015MUSANAM}
D.~Snyder, G.~Chen, and D.~Povey, ``Musan: A music, speech, and noise corpus,''
  \emph{ArXiv}, vol. abs/1510.08484, 2015.

\bibitem{ko2017rir}
T.~Ko, V.~Peddinti, D.~Povey, M.~L. Seltzer, and S.~Khudanpur, ``A study on
  data augmentation of reverberant speech for robust speech recognition,'' in
  \emph{2017 IEEE International Conference on Acoustics, Speech and Signal
  Processing (ICASSP)}, 2017, pp. 5220--5224.

\bibitem{park2019specaug}
\BIBentryALTinterwordspacing
D.~S. Park, W.~Chan, Y.~Zhang, C.-C. Chiu, B.~Zoph, E.~D. Cubuk, and Q.~V. Le,
  ``Specaugment: A simple data augmentation method for automatic speech
  recognition,'' \emph{Interspeech 2019}, Sep 2019. [Online]. Available:
  \url{http://dx.doi.org/10.21437/Interspeech.2019-2680}
\BIBentrySTDinterwordspacing

\bibitem{deng2019aamsoftmax}
J.~Deng, J.~Guo, N.~Xue, and S.~Zafeiriou, ``Arcface: Additive angular margin
  loss for deep face recognition,'' in \emph{2019 IEEE/CVF Conference on
  Computer Vision and Pattern Recognition (CVPR)}, 2019, pp. 4685--4694.

\end{thebibliography}

\end{document}